\documentclass[twocolumn]{aastex631}
\usepackage{graphicx} 
\usepackage{epstopdf}
\usepackage{amsmath,soul} 
\usepackage{placeins}

\newcommand\rmd{\mathrm{d}}
\newcommand\ompi{\omega_{\mathrm{p},i}}

\newcommand\be{\begin{equation}}
\newcommand\ee{\end{equation}}

\shorttitle{Ion--Electron Relativistic Reconnection}
\shortauthors{F.\ Bacchini et al.}

\begin{document}

\title{Three-dimensional Dynamics of Strongly Magnetized Ion--Electron Relativistic Reconnection}

\author[0000-0002-7526-8154]{Fabio Bacchini}
\affiliation{Centre for mathematical Plasma Astrophysics, Department of Mathematics,
KU Leuven, Celestijnenlaan 200B, B-3001 Leuven, Belgium}
\affiliation{Royal Belgian Institute for Space Aeronomy, Solar-Terrestrial Centre of Excellence, Ringlaan 3, 1180 Uccle, Belgium}
\author[0000-0001-9039-9032]{Gregory R.\ Werner}
\affiliation{Center for Integrated Plasma Studies, Department of Physics, University of Colorado, 390 UCB, Boulder, CO 80309-0390, USA}
\author[0000-0003-2841-8153]{Camille Granier}
\affiliation{Centre for mathematical Plasma Astrophysics, Department of Mathematics,
KU Leuven, Celestijnenlaan 200B, B-3001 Leuven, Belgium}
\author[0000-0003-3349-7394]{Jesse Vos}
\affiliation{Centre for mathematical Plasma Astrophysics, Department of Mathematics,
KU Leuven, Celestijnenlaan 200B, B-3001 Leuven, Belgium}

\correspondingauthor{Fabio Bacchini}
\email{fabio.bacchini@kuleuven.be}

\begin{abstract}
We present 3D simulations of semirelativistic collisionless magnetic reconnection, where upstream ions are subrelativistic while electrons are ultrarelativistic. We employ the realistic proton-to-electron mass ratio and explore a range of upstream ion magnetization spanning two orders of magnitude, with our highest-magnetization run achieving unprecedentedly large domain sizes. Through a parameter scan, we find that as the system transitions from mildly to trans- and ultrarelativistic regimes the qualitative behavior of reconnection becomes strongly influenced by 3D effects mediated by drift-kink and flux-rope kink dynamics. As a result, magnetic-energy dissipation at high magnetizations, and the subsequent nonthermal particle acceleration, can become less efficient, contrary to general expectations for 3D relativistic reconnection. Our results have important implications for understanding reconnection in magnetized astrophysical scenarios, such as the surroundings of black holes and neutron stars.
\end{abstract}

\section{Introduction}
\label{sec:intro}
Many astrophysical systems dissipate magnetic energy through plasma processes with important consequences for their dynamical evolution and/or observational signatures.  
However, how that energy is ultimately released at kinetic (particle) scales, which are many orders of magnitude smaller than the global system size, remains an open question relevant for various plasma environments around astrophysical sources, such as the surroundings of compact objects (e.g.\ black holes and neutron stars). 
An absolutely necessary plasma-astrophysical endeavor for these physical scenarios is to understand how one of these processes, magnetic reconnection, which occurs at current sheets where magnetic-field reversals arise, converts magnetic energy into plasma heating and nonthermal particle acceleration.
This paper uses Particle-in-Cell (PIC) simulation to explore collisionless magnetic reconnection and particle energization in the semirelativistic regime where electrons are relativistic but ions (with similar temperature to electrons, and realistic mass ratio) are subrelativistic.  Furthermore, it explores this regime in both 2D and 3D, thanks in part to the recent development of a semiimplicit algorithm with superior stability properties.

Reconnection in semirelativistic ion--electron plasmas has so far received less attention than both the nonrelativistic and (ultra)relativistic regimes, although understanding the relative energization of electrons and ions is of critical importance for global magnetohydrodynamic (MHD) modeling of radiation from black-hole accretion flows \citep[e.g.][]{chael2018,dexter2020a,ressler2020,hankla2022,scepi2022,ressler2023,vos2024}.
Unlike in the (ultra)relativistic regime where ions behave essentially as positrons yielding symmetric behavior between electrons and ions \citep[unless radiative effects are considered; e.g.][]{chernoglazov2023}, in the semirelativistic regime electrons and ions at the same temperature have different gyroradii and behave differently as they are energized.
Important progress has been made through 2D PIC studies of reconnection in this regime \citep{melzani2014a,melzani2014b,guo2016pop,rowan2017,ball2018pic,werner2018,ball2019,rowan2019}.  However, there remains much we do not understand about reconnection and other dissipation mechanisms in 3D current sheets.
In relativistic electron--positron plasma, some earlier PIC studies concluded that the 3D behavior of initially stable current sheets is fairly similar to the 2D dynamics \citep{yin2008,liu2011,kagan2013,guo2014,sironi2014,guo2015,werneruzdensky2017,guo2021}, although there is significant evidence that, at the very least, particle energy in 2D is limited by the difficulty of escape from plasmoids formed by reconnection, whereas in 3D particles escape and undergo additional acceleration \citep[e.g.][]{dahlin2015,petropoulou2018,li2019,hakobyan2021,zhangH2021,zhangH2023}.
In 2D collisionless reconnection, dynamics is restricted to the plane where tearing modes can develop and grow. Plasmoids can form from reconnection, and in periodic or closed domains, they accumulate and create system-size-scale structures where particles are trapped indefinitely. This slows down (or halts) reconnection, preventing further dissipation of magnetic energy. In 3D domains, additional out-of-plane modes such as the drift-kink instability (DKI) and the flux-rope (MHD) kink instability may develop, folding and rippling current sheets perpendicularly to the reconnection plane. The DKI in particular is stronger for higher magnetizations and for plasmas with small scale separation between species, (e.g.\ pair plasma, or the semirelativistic ion--electron plasma considered in this work), and may nonlinearly interact with tearing modes. MHD kinking, instead, can destroy elongated magnetic structures (plasmoids in 2D) created by reconnection, liberating trapped particles and contributing to their energization.
In some cases, these kinking modes can alter the mechanism and nature of magnetic energy dissipation and particle energization \citep[e.g.][]{pritchettetal1996,zenitanihoshino2005apjl,zenitanihoshino2007,zenitanihoshino2008,hoshino2020,werneruzdensky2021,hoshino2024}.
These differences have been found to carry over to the 3D semirelativistic regime \citep{werneruzdensky2024}, but very limited exploration of 3D dynamics has been carried out so far. In particular, it is unclear how previous results (focusing on low magnetization) extend to strongly magnetized regimes. 

In this Letter, we compare 2D and 3D PIC simulations of current sheets exhibiting magnetic reconnection for an unpredecentedly wide range of magnetizations, spanning two orders of magnitude and exploring increasingly larger physical domain sizes. Our highest-magnetization run thus achieves the status of largest ion--electron reconnection simulation to date. We focus in particular on the competition of instabilities determining the overall current-sheet evolution in 3D, and on how the instability interplay changes as the magnetization varies.

This Letter is organized as follows: In Section~\ref{sec:setup}, we present our numerical setup and the details of the various simulations performed. In Section~\ref{sec:results}, we discuss our main results on magnetic-energy dissipation, 3D current-sheet dynamics and structure, and particle acceleration. Finally, in Section~\ref{sec:conclusion} we discuss our results and present our conclusions.

\section{Numerical Setup}
\label{sec:setup}
We perform PIC simulations of ion--electron semirelativistic plasma with the \textsc{RelSIM} code (\citealt{bacchini2019fipic,bacchini2023}). Our setups are generally similar to \cite{werneruzdensky2024}: a Harris equilibrium is initialized in a double-periodic simulation domain $(x,y,z)\in[L_x\times L_y\times L_z]$ with $L_x=L_y/2=L_z=L$, such that we keep a fixed ratio $L/(\sigma_{i,0} \rho_{i,0})\simeq 55$, where $\sigma_{i,0} \equiv B_0^2/(4\pi m_i n_0 c^2)$ is the upstream (cold) ion magnetization and $\rho_{i,0}\equiv m_ic/(q_i B_0)$ is a measure of the upstream ion Larmor radius\footnote{This choice is guided by the expectation that, at steady state, reconnection energizes upstream particles such that their Larmor radius increases from roughly $\rho_{i,0}$ to $\sigma_{i,0}\rho_{i,0}$. However, this only applies in relativistic cases $\sigma_{i,0}\gg1$, and for the low-$\sigma_{i,0}$ cases analyzed in this work the actual energy scaling may deviate from this expectation.}. Here, $B_0$ is the upstream magnetic field, $m_i$ is the ion mass, and $n_0=n_{i,0}=n_{e,0}$ is the upstream particle number density, with 8 particles per cell per species. The upstream plasma is initialized as a Maxwellian for both particle species, with ions having (dimensionless) temperature $\theta_{i,0}\equiv kT_{i,0}/(m_ic^2)=0.01$. For the electrons, we set $T_{e,0}/T_{i,0}=1$; we employ the realistic mass ratio $m_i/m_e=1836$, resulting in a dimensionless initial electron temperature $\theta_{e,0}\equiv kT_{e,0}/(m_ec^2)=18.36$ for all runs. The initial average electron Lorentz factor in the upstream is thus $\gamma_{e,0}\sim 55$ and the plasma is effectively semirelativistic.

We explore a range of magnetizations $\sigma_{i,0}\in[0.1,10]$ by varying $B_0$ and keeping all other parameters fixed. Our highest-magnetization run with $\sigma_{i,0}=10$ employs a system size of $180\times360\times180 (c/\ompi)^3$ and is the largest 3D ion--electron simulation conducted to date. In all cases, the numerical resolution is such that the grid spacing is $\Delta x=\Delta y=\Delta z \simeq 0.15 c/\ompi$ and the time step is $\Delta t = 0.1\ompi^{-1}$, where $\ompi\equiv\sqrt{4\pi q_i^2 n_0/m_i}$. All simulations are run until $t_\mathrm{f}=2000\ompi^{-1}$. For our magnetization scan, this results in $t_\mathrm{f} \in [11,111]L/c$ (from highest to lowest value of $\sigma_{i,0}$). In terms of electron scales, the spatiotemporal resolutions are $\Delta x \simeq c/\omega_{\mathrm{p},e}$ and $\Delta t \simeq 0.6\omega_{\mathrm{p},e}^{-1}$, where $\omega_{\mathrm{p},e}\equiv\sqrt{4\pi q_e^2 n_0/(\gamma_{e,0}m_e)}$ is the relativistic electron plasma frequency. Note that this resolution is roughly 7 times smaller (in each spatial direction) than the resolution employed by \cite{werneruzdensky2024}\footnote{Cf.\ the simulation presented in \cite{werneruzdensky2024} ($\sigma_{i,0}=0.5$), which we reproduce here with almost exactly the same parameters but 7 times smaller resolution in each direction, obtaining equivalent results.}. This is made possible by \textsc{RelSIM}'s implicit approach, which ensures much stronger numerical stability at coarser numerical resolutions with respect to traditional explicit PIC (\citealt{bacchini2023}).

The initial Harris equilibrium is constructed without any guide field and with the standard magnetic-field profile
\be
B_{x,0}(y) = B_0 \left[1 - \tanh\left(\frac{y-L_y/4}{\delta_0}\right) +\tanh\left(\frac{y-3L_y/4}{\delta_0}\right)\right],
\ee
where $\delta_0=0.5c/\ompi$ is the initial current-sheet half-thickness. A hot, drifting ion--electron population localized in the current sheet provides the pressure support at equilibrium; its number density $n_\mathrm{CS}$ is such that an overdensity $\eta=n_\mathrm{CS}/n_0=5$ is achieved in the center of the current sheet. To isolate the pure competition of tearing vs.\ kink modes without imposed biases, no perturbation is applied to the initial equilibrium, such that the system's initial evolution is driven by numerical noise (in Section~\ref{sec:conclusion} we discuss the possible implications of this choice).

In the following, we also present 2D simulations for direct comparison. These are initialized exactly as described above for 3D runs, except that variations along the $z$-direction are ignored.

\begin{figure*}
    \centering
    \includegraphics[width=1\linewidth]{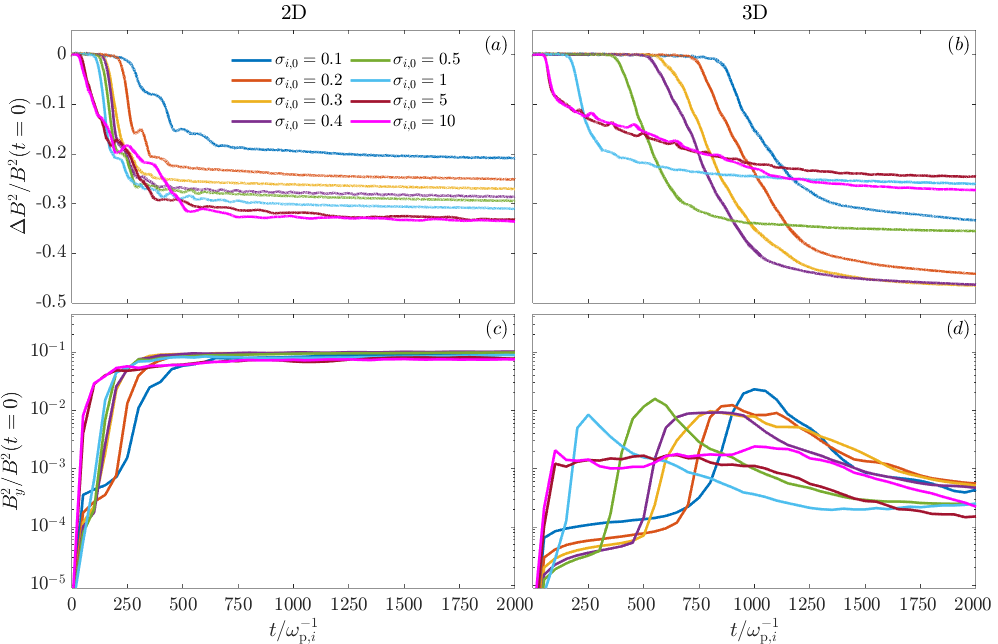}
    \caption{Top row: Magnetic-energy dissipation over time in 2D (panel $a$) and 3D (panel $b$) during ion--electron reconnection for different $\sigma_{i,0}$. Bottom row: Evolution of magnetic energy stored in the (initially absent) $B_y$ component in 2D (panel $c$) and 3D (panel $d$).
    }
    \label{fig:enB_sigmacomp}
\end{figure*}

\begin{figure*}
    \centering
    \includegraphics[width=1\linewidth]{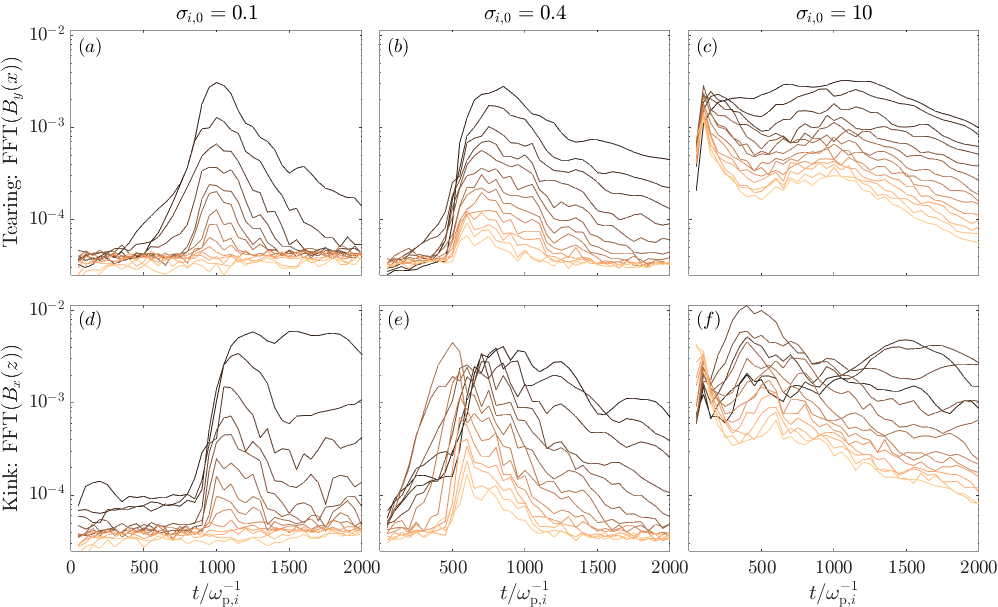}
    \caption{Evolution of the spectral power in selected wavenumbers of $B_y(x)$ (top row) and $B_x(z)$ (bottom row) at the location of the current sheet, for $\sigma_{i,0}=0.1,0.4,10$ (from left to right columns). Each line corresponds to a different value of $k$, with darker lines representing smaller $k$.
    }
    \label{fig:fft_sigmacomp}
\end{figure*}

\section{Results}
\label{sec:results}

\subsection{Magnetic-energy Dissipation}
\label{sec:dissipation}

In our runs, we can generally identify an initial ``quiet'' phase during which the initial equilibrium remains marginally stable and (noise-induced) perturbations start growing. Over time, these fluctuations reach amplitudes large enough to induce reconnection and fragment the initial current sheet. At that point, magnetic energy is violently released through an ``active'' reconnection phase; a subsequent, final phase is then initiated, whose dynamics depends on upstream conditions, as we discuss below. In general, during active reconnection, we can identify the presence of flux tubes extending along $z$ and persistent chaoticity in the current-sheet structure. Toward the end of the simulation, the current sheet becomes thick and rather unstructured, without visible macroscopic flux tubes. This dynamics qualitatively agrees with the results of \cite{werneruzdensky2021,werneruzdensky2024}. {As described in Section~\ref{sec:setup}, all our simulations are run for the same physical time $t_\mathrm{f}=2000\ompi^{-1}$, which for low magnetization (therefore small $L$) is equivalent to $\sim111L/c$. A priori, it would seem unnecessary to let the system evolve over so many light-crossing times, since with the standard (collisionless) reconnection rate in the relativistic limit ($\sim 0.1c$) all magnetic energy would be depleted much faster. However, at low magnetization the reconnection rate will be far from the relativistic expectation, and therefore it becomes necessary to run for longer times.}

Fig.~\ref{fig:enB_sigmacomp} (top row) shows the evolution in time of the mean dissipated magnetic energy, normalized to the total initial magnetic energy. In particular, we compare the results of our 3D runs (panel $b$) with equivalent 2D runs (panel $a$). From this analysis, we observe remarkable differences due to out-of-plane dynamics.

First, we note that 2D runs display a very clear trend: higher $\sigma_{i,0}$ results in more magnetic-energy dissipation, up to a fraction of $\sim30\%$ of the initial available energy for the highest $\sigma_{i,0}=10$ at the end of the run ($t=2000\ompi^{-1}$). The close similarity between the two highest $\sigma_{i,0}$ suggests that the evolution becomes independent of the magnetization (and therefore system size, which increases with $\sigma_{i,0}$), and that our results are converged. For 3D runs, the trend is strikingly nonmonotonic: energy dissipation increases from $\sigma_{i,0}=0.1$ to 0.4, reaching a maximum of $\sim45\%$ of the initial magnetic energy. For $\sigma_{i,0}\ge0.5$, the trend reverses and \emph{less} energy is dissipated for higher magnetizations. The amount of energy dissipated converges to $\sim25\%$ of the initially available energy at our highest $\sigma_{i,0}=10$ (consistent with $\sigma_{i,0}=5$, which again indicates convergence).

A second difference is that for $\sigma_{i,0}\le0.5$, 3D runs dissipate more energy than corresponding 2D cases, in accordance with \cite{werneruzdensky2024} and general expectations. However, above this threshold, we find that 3D cases dissipate \emph{less} energy than corresponding 2D runs.

Finally, we note how magnetic-energy dissipation proceeds differently, during different phases of the evolution, for different $\sigma_{i,0}$. In 2D, there exists a clear trend in which most of the energy is dissipated during an active phase that proceeds rapidly and at the end of which very little further dissipation occurs. In 3D, the active reconnection phase may proceed rapidly and dissipate most of the energy (for low $\sigma_{i,0}$), mimicking 2D behavior, or conversely begin with the rapid release of a fraction of magnetic energy and then progressively slow down (for high $\sigma_{i,0}$).

These findings contradict the general expectation that, in 3D, reconnection dissipates more energy than in 2D by means of out-of-plane dynamics (e.g. \citealt{werneruzdensky2021,zhangQ2021,werneruzdensky2024}). 3D dynamics does take place, but its efficiency in liberating more magnetic energy appears to depend on the magnetization, with high $\sigma_{i,0}$ values introducing qualitative changes that deserve exploration. As we discuss below, a possible explanation lies in the competition between tearing, DKI, and MHD-kink instabilities driving reconnection dynamics.

\subsection{Tearing versus Kink Modes in 3D}

As discussed in Section~\ref{sec:intro}, when increasing $\sigma_{i,0}$, the dynamics of and interplay between tearing, DKI, and MHD-kink modes may qualitatively change, thereby changing the overall reconnection properties. To investigate this possibility, we diagnose the development of tearing and kinking modes.

First, as a qualitative indicator of magnetic-topology rearrangement we monitor the creation of a $B_y$ component, which is initially absent in our setup. As shown in Fig.~\ref{fig:enB_sigmacomp} (bottom row), in 2D (panel $c$) the creation of $B_y$ follows from active reconnection, where some of the energy in the reconnecting field $B_x$ is converted into perpendicular $B_y$ energy. This reaches around $10\%$ of the initial magnetic energy (independently of $\sigma_{i,0}$) and then plateaus, since no further significant destruction of magnetic fields can occur after active reconnection has ended. In 3D (panel $d$), instead, energy in $B_y$ generally grows at a slower rate than in 2D and also reaches a much smaller peak value; after that, $B_y$ energy is slowly depleted. We observe that this behavior is $\sigma_{i,0}$-dependent: Lower magnetizations correspond to slower growth and larger peak values of $B_y$ energy.
This observed 3D dynamics is compatible with the development of DKI and MHD-kink modes: DKI may slow down and impede tearing-mediated reconnection, especially at high $\sigma_{i,0}$, whereas MHD kink may deplete $B_y$ energy after (and if) tearing creates elongated flux tubes along $z$.
The strikingly different way in which 2D and 3D reconnection proceeds thus suggests that out-of-plane kink modes could be playing a determining role in realistic (i.e.\ 3D) situations, heavily affecting the overall reconnection process.

To quantitatively assess the growth of tearing and kink modes, we compute the Fourier spectrum of magnetic-field fluctuations in the current-sheet plane. This is shown for selected wavenumbers in Fig.~\ref{fig:fft_sigmacomp} for $B_y(x)$ (top row, indicating tearing) and $B_x(z)$ (bottom row, indicating kink activity) for low ($\sigma_{i,0}=0.1$, left column), intermediate ($\sigma_{i,0}=0.4$, middle column), and high ($\sigma_{i,0}=10$, right column) magnetizations. The intermediate $\sigma_{i,0}=0.4$ case was selected as a representative of the threshold magnetization above which the energy-dissipation trend in 3D reverses (cf.\ Fig.~\ref{fig:enB_sigmacomp}$b$). {Here, we plot the first 15 wavenumbers $k=[k_\mathrm{min}, 2k_\mathrm{min}, 3k_\mathrm{min}, ...]$}, with darker lines corresponding to smaller wavenumbers and with $k_\mathrm{min}\equiv\pi/L$.
Here, we focus on two key parameters: the time scales of the growth of tearing and kink modes, and the power in different modes.

At low $\sigma_{i,0}$ (panels $a$ and $d$), tearing modes begin to develop much faster than kink modes\footnote{{Note that panel $d$ shows the presence of kink oscillations at very early times at powers (a factor few) larger than those at which tearing modes exist. However, this power is associated with small $k$, i.e.\ with long wavelengths, which normally would indicate MHD-kinking. At this early stage, however, MHD-kink cannot be developing yet, and therefore we interpret this signal as some initial oscillation of the whole sheet, potentially induced by the initial conditions, without a specific physical meaning.}}, implying that the current sheet experiences thinning and breaking in a 2D-like fashion, before significant kink fluctuations develop. Power in tearing modes peaks, and shortly after that kink modes reach saturation. Tearing modes then start decreasing in power, while kink modes remain active. It is noteworthy that the only significant kink modes are those at the smallest $k$ (i.e.\ at wavelengths comparable to the system size), implying that MHD-kink is active but DKI is not. This aligns with linear theory, which predicts that DKI is rather insignificant at low magnetization (e.g.\ \citealt{zenitanihoshino2005prl}).

At intermediate $\sigma_{i,0}$ (panels $b$ and $e$), tearing and kink modes grow over comparable time scales. Here, DKI modes (identified by larger wavenumbers) start growing slightly faster than tearing; MHD-kink modes (at smaller $k$) instead grow slightly later. As a result, the current sheet first becomes ``wavy'' and folded due to DKI before tearing can break up the current layer, and after that the MHD-kink can grow and destroy the elongated flux tubes. After a peak in power, tearing modes become weaker, albeit at a slower rate with respect to the $\sigma_{i,0}=0.1$ case. Note that, as in the previous case, power in tearing and kink modes saturates around the same value.

At high $\sigma_{i,0}$ (panels $c$ and $f$), the dynamics is significantly different. First, power in small-$k$ kink modes peaks in the very beginning of the run and then rapidly decreases, indicating a fast activation of DKI. Tearing modes start growing immediately after, but take much longer to saturate. Longer-wavelength kink modes grow slightly slower than tearing, but saturate at substantially ($\sim$3--4 times) larger power. In addition, the longest-wavelength MHD-kink modes grow very slowly and only reach saturation around the end of the run, with the lowest-$k$ mode reaching significantly smaller peak power than other tearing and kink modes. Therefore, the current layer effectively undergoes very strong DKI very quickly, potentially delaying the saturation of tearing, and at the same time MHD-kink modes arise only very late in the run if at all.

This analysis shows that larger $\sigma_{i,0}$ values are strongly correlated with a qualitative change in unstable-mode time scales and subsequent magnetic-energy dissipation dynamics. At low magnetization, DKI is absent; tearing creates flux ropes which can decay via MHD kink. At high magnetization, DKI is intensely active, tearing is slowed down, and long flux ropes form less efficiently, therefore impeding MHD-kink. As also discussed in \cite{werneruzdensky2021}, this implies that magnetic-energy dissipation in 3D strongly depends on the kink--tearing interplay and involves a modification in the current-sheet structure. We analyze and discuss this aspect in the next section.

\begin{figure*}
    \centering
    \includegraphics[width=1\linewidth, trim={10cm 0 10cm 0}, clip]{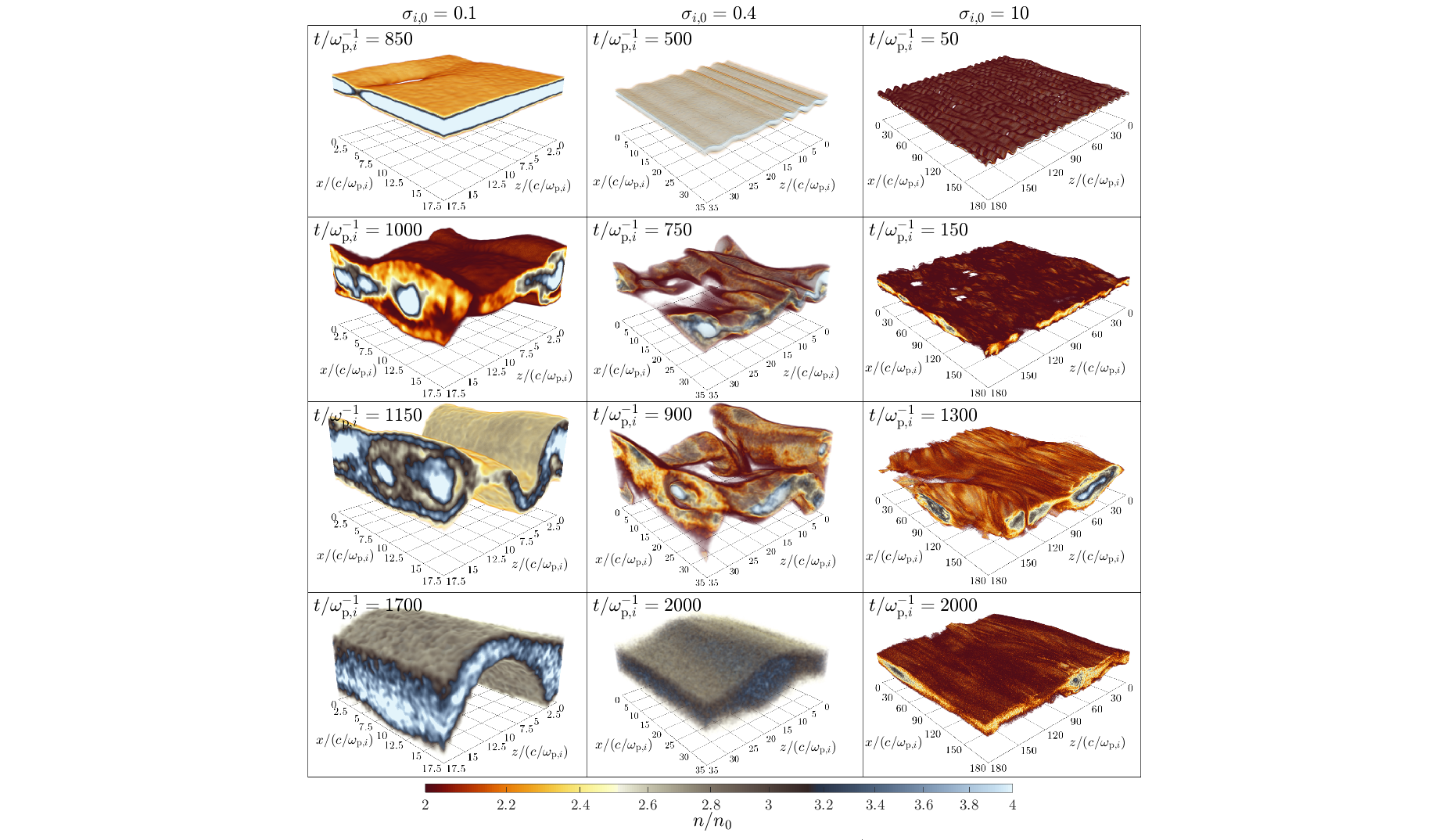}
    \caption{Volume rendering {(covering the entire simulation domain)} of the total (ion and electron) plasma density in the current sheet during subsequent reconnection phases (from top to bottom) and for increasing magnetization (from left to right).}
    \label{fig:csstructure}
\end{figure*}

\subsection{Current-sheet Structure and Kinking Flux Tubes}

The analyses presented in the previous sections indicate that magnetic-energy dissipation is strongly influenced by the interplay of tearing and kink modes, which changes depending on the magnetization. The observed result is that, at low $\sigma_{i,0}$ more magnetic energy is dissipated in 3D than in 2D (cf.\ Fig.~\ref{fig:enB_sigmacomp}), and the opposite occurs at high $\sigma_{i,0}$. As shown above, reconnection-driven flux-rope creation and kink-driven destruction may be impeded at high $\sigma_{i,0}$, which could explain the difference in magnetic-energy dissipation, as also discussed in previous works (e.g.\ \citealt{werneruzdensky2021,zhangH2021,werneruzdensky2024}).

To confirm this reasoning, in Fig.~\ref{fig:csstructure} we show representative snapshots of the plasma density in the current sheet at different times, for $\sigma_{i,0}=0.1$ (top row), 0.4 (middle row), and 10 (bottom row).
At $\sigma_{i,0}=0.1$, the temporal evolution of the current sheet first shows the occurrence of tearing, followed by the creation of flux ropes, their violent MHD-kinking and breaking, and a final stage in which the current sheet assumes a thick and chaotic structure. No evident DKI fluctuations are present during the initial stages of the evolution, where tearing essentially proceeds 2D-like.
At $\sigma_{i,0}=0.4$, DKI fluctuations appear first, creating a wavy current sheet but without compromising the sheet's overall structure. Tearing then follows, creating flux ropes which then kink and dissipate. The sheet becomes thick and chaotic in the final stages of evolution. At $\sigma_{i,0}=10$, strong DKI fluctuations immediately appear, and the sheet starts to thicken from the very beginning of the run. In this situation, tearing, and the ensuing creation of flux ropes, is strongly impeded because the sheet thickness remains large due to the action of the DKI. Reconnection proceeds differently, with the sheet assuming an increasingly thicker and more chaotic structure and without elongated flux ropes appearing and violently kinking.

\begin{figure}
    \centering
    \includegraphics[width=1\linewidth]{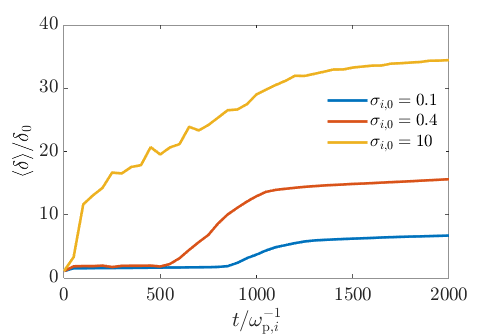}
    \caption{Evolution of the average current-sheet thickness $\delta$, defined as the region where $\sigma_i\le3\sigma_{i,0}/4$, for selected initial magnetizations.}
    \label{fig:csthickness}
\end{figure}

These results indicate that DKI-driven sheet thickening, which occurs faster and more strongly at high $\sigma_{i,0}$, is the key mechanism driving qualitatively different reconnection dynamics in our 3D simulations. This is further supported by Fig.~\ref{fig:csthickness}, where we show the evolution of the average current-sheet thickness $\delta$, defined as the region where $\sigma_i\le3\sigma_{i,0}/4$, i.e.\ where the magnetization drops by at least 25\% from its upstream value\footnote{{We have tested different thresholds for this analysis down to $\sigma_i\le0.5\sigma_{i,0}$, obtaining qualitatively similar results.}}. We observe that the sheet becomes much thicker at a much faster rate for higher magnetizations. This again suggests that DKI dynamics changes the sheet structure at high $\sigma_{i,0}$, maintaining the sheet in a chaotic, increasingly thicker state in which tearing modes (and the creation of large-scale flux ropes) are strongly suppressed.

\subsection{Particle acceleration}

\begin{figure*}
    \centering
    \includegraphics[width=1\linewidth]{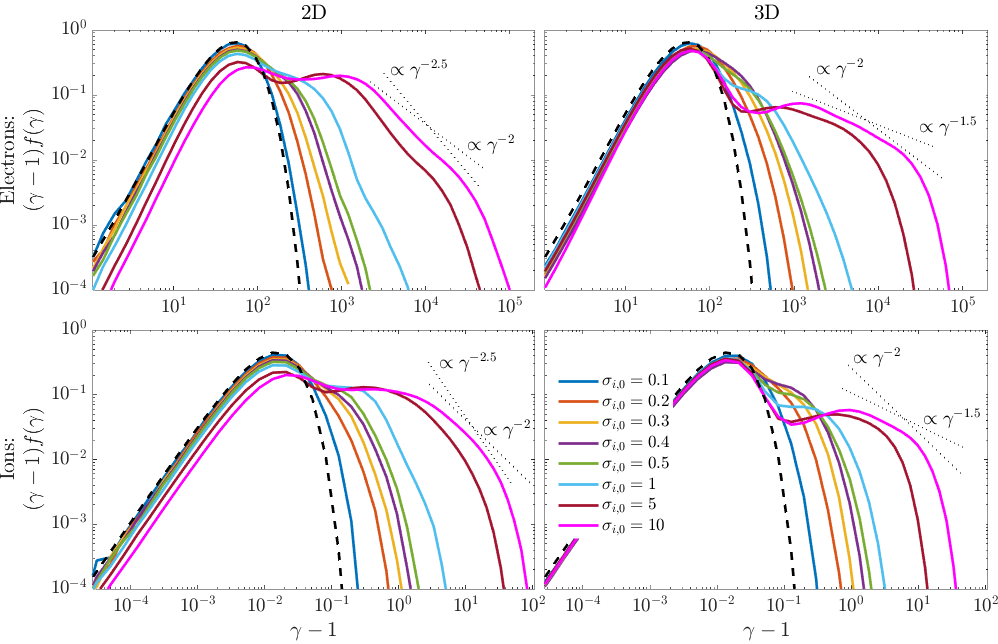}
    \caption{Energy distributions of electrons (top row) and ions (bottom row) at the end of each run ($t=2000\ompi^{-1}$) in 2D (left column) and 3D (right column). Dashed lines indicate the initial distribution. Dotted lines show reference spectral slopes.}
    \label{fig:dist_sigmacomp}
\end{figure*}

Our ion--electron simulations allow us to investigate how different plasma species can gain energy via reconnection. To conclude our analysis, Fig.~\ref{fig:dist_sigmacomp} shows the energy distributions $f(\gamma)\equiv\rmd\mathrm{N}/\rmd\gamma$ {(excluding the initial drifting populations)} of electrons (top row) and ions (bottom row) for equivalent 2D (left column) and 3D (right column) simulations, for $\sigma_{i,0}\in[0.1,10]$ and at the end of each run ($t_\mathrm{f}=2000\ompi^{-1}$). The initial distribution at $t=0$ is shown as a dashed black line. The trend we retrieve corresponds to the general expectation that higher $\sigma_{i,0}$ produce higher-energy particles (e.g.\ \citealt{werneruzdensky2021}), but with important differences between species and between 2D and 3D runs.

First, we note that both particle species can develop nonthermal populations, with $\sigma_{i,0}=10$ producing evident nonthermal tails extending for up to 2 decades in both 2D and 3D. However, the slope of the nonthermal tail differs depending on dimensionality, with 3D runs producing markedly less steep tails. Reference slopes of $\gamma^{-1.5}$, $\gamma^{-2}$, and $\gamma^{-2.5}$ are shown to guide the eye; this broadly aligns with previous findings for 3D reconnection (\citealt{zhangH2021,zhangH2023}). In these runs, ions are initially nonrelativistic and, for the highest $\sigma_{i,0}$, can reach maximum Lorentz factors of $\sim10$--$100$; electrons are already relativistic at initialization, and their energy gain is such that $\gamma$ can increase by over 2 orders of magnitude (up to $\sim10^5$) for the case with the highest magnetization. The maximum Lorentz factor is roughly comparable in 2D and 3D for both species, but in 3D it is generally smaller by a factor 2--3 for the highest $\sigma_{i,0}$.

\begin{figure}
    \centering
    \includegraphics[width=1\linewidth]{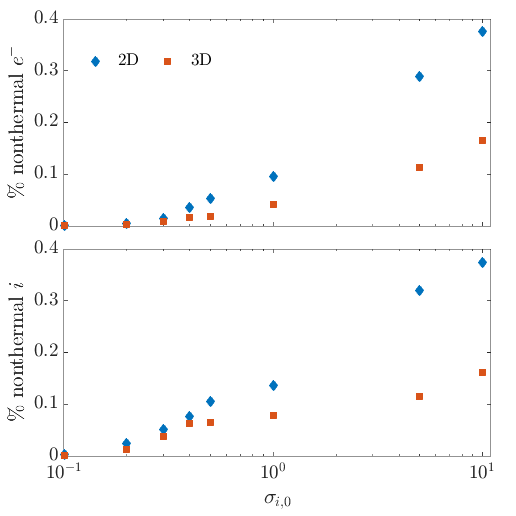}
    \caption{Fraction of nonthermally accelerated electrons (top row) and ions (bottom row) at the end of each run ($t=2000\ompi^{-1}$), for different magnetizations in 2D (diamonds) and 3D (squares). This is obtained by fitting a Maxwell-J\"uttner distribution to each spectrum and then subtracting the thermal part.}
    \label{fig:accfrac}
\end{figure}

In addition, we observe that, at high $\sigma_{i,0}$, the nonthermal part of the spectrum appears to contain fewer particles in 3D than in 2D. To quantify this difference, at the end of each run ($t=2000\ompi^{-1}$) we count the fraction of nonthermal particles by fitting a Maxwell--J\"uttner distribution to each spectrum and subtracting the corresponding thermal part. The result is shown in Fig.~\ref{fig:accfrac}: we observe that 3D runs, especially at high $\sigma_{i,0}$, systematically accelerate a (up to twice) smaller fraction of particles toward nonthermal energies.

These results align with our finding (see Section~\ref{sec:dissipation}) that 3D reconnection at high magnetization may dissipate less magnetic energy than in 2D, thereby achieving lower acceleration efficiency. This outcome has potentially profound consequences on our understanding of magnetic-field dynamics in compact-object environments, as we discuss in greater detail below.

\section{Discussion}
\label{sec:conclusion}

In this Letter, we have reported on a series of fully kinetic simulations of collisionless magnetic reconnection in semirelativistic ion--electron plasma with realistic mass ratio $m_i/m_e=1836$. We have explored a range of upstream ion magnetizations $\sigma_{i,0}\in[0.1,10]$ covering two orders of magnitude. Our largest run with $\sigma_{i,0}=10$ employs a system size of $180\times360\times180 (c/\ompi)^3$, achieving the status of largest 3D ion--electron simulation conducted to date. This allows us to explore how reconnection dynamics changes from nonrelativistic to trans- and ultrarelativistic scenarios over an unprecedented range of parameters, with the highest-magnetization cases being relevant for compact-object magnetospheres where ion--electron plasma may be subjected to extremely strong magnetic fields (e.g.\ \citealt{yuannarayan2014,blandford2019,guo2024} and references therein).
While in the ultrarelativistic limit $\sigma_{i,0}\to\infty$ we expect ion--electron reconnection to behave equivalently to the more thoroughly explored pair-plasma case (e.g.\ \citealt{guo2016apjl,ball2018,werner2018,petropoulou2019}), our two-species setup still offers insight into each species' dynamics separately, particularly in terms of energization. 

We find that low-$\sigma_{i,0}$ runs proceed initially in a 2D-like fashion; tearing modes grow first, creating elongated flux tubes (plasmoids in 2D) which can be destroyed by violent MHD kinking. Short-wavelength DKI modes are relatively unimportant, and overall this results in the dissipation of more magnetic energy than in 2D. At moderate $\sigma_{i,0}=0.4$, DKI modes become strong enough to drive additional magnetic-energy destruction without impeding tearing modes. The latter are still free to develop and break the sheet, creating large-scale MHD-kink-prone flux ropes. Essentially, up until $\sigma_{i,0}=0.4$ we observe no strong competition between tearing and DKI modes, and their combined action releases a much larger fraction of magnetic energy in 3D with respect to 2D (about 45\% versus 27\%, see Fig.~\ref{fig:enB_sigmacomp}$a$--$b$). In high-$\sigma_{i,0}$ ($\ge5$) runs, instead, DKI grows rapidly and actively impedes the thinning of the sheet and the development of tearing. Large-scale flux ropes develop less freely, and instead the current sheet quickly becomes a thick and chaotic structure where magnetic fields are destroyed much more slowly{\footnote{While \cite{zhangH2021} reported an observed slowdown of pair-plasma reconnection in 3D vs.\ 2D, the slowdown observed in our simulations is much more significant.}}. As a result, a smaller fraction of magnetic energy is dissipated in 3D than in 2D (27\% versus 33\%). In this situation, particle acceleration is roughly twice less efficient in 3D than in equivalent 2D runs where kinking is absent.

{While several points of what reported above deserve more exploration (for example, it remains unclear why $\sigma_{i,0}=0.4$ represents a threshold for the inversion in the energy-dissipation trend),} these findings contradict the standard picture in which 3D reconnection dissipates more magnetic energy than in 2D (e.g. \citealt{werneruzdensky2021,zhangQ2021,werneruzdensky2024}). In particular, our results describe a scenario in which reconnecting plasma gathers in slowly dissipating current sheets, which are kept thick and chaotic by DKI. Such a very thick current sheet may even be described as two separate sheets, where upstream, cold- and magnetized-plasma conditions meet an internal hot, lowly magnetized plasma state. In this situation, standard expectations for symmetric Harris-sheet reconnection may not be applicable. Our high-$\sigma_{i,0}$ results are especially relevant for astrophysical scenarios such as compact-object magnetospheres and coronae, where ultramagnetized plasma could be undergoing reconnection, potentially powering high-energy flares. This may have implications for comparing the time scales of reconnection with flaring activity observed from high-energy sources, although quantifying the impact of the current-sheet structure on flare dynamics requires dedicated efforts to be pursued elsewhere.

While our simulations paint a novel picture of ion--electron reconnection in highly magnetized plasma, it is necessary to discuss their generality. At the time of writing, there exist no exactly equivalent studies (nor theory) to which our work can be compared directly; but the thorough analysis of 3D pair-plasma reconnection (in closed domains) presented in \cite{werneruzdensky2021} is a good starting point. In one of their simulations, they consider a high-magnetization case, $\sigma_0=25$, and find results very similar to ours: a slowdown in time of magnetic-energy dissipation and less efficient particle acceleration. Notably, they report that this is observed only when \emph{no guide field and no initial perturbation are employed.} \cite{werneruzdensky2021} also conjecture on why a guide field and/or an initial, localized perturbation would result in a different reconnection dynamics: in practice, they point to the presence of initially drifting plasma species in the current sheet (which provide the initial equilibrium) as responsible for changing reconnection dynamics at large $\sigma_0$. The drifting species are initialized differently for different $\sigma_0$, and for high magnetization in particular, the current-sheet plasma at initialization is much hotter than the background (and thus possesses significant inertia). If not swept away and confined within regions where it cannot further affect the current-sheet dynamics, the initial plasma could play a role in determining reconnection efficiency. \cite{werneruzdensky2021} propose that guide fields or initial perturbations may act in the direction of confining this hot plasma and preventing effects such as slowdown of magnetic dissipation. However, whether this confinement is a physical effect or is, in turn, an intrinsic feature of limited-size simulations remains to be established. We also emphasize that our use of periodic boundaries may have an effect on the results, since it constrains hot reconnected plasma to remain inside the box. However, periodic boundaries also allow us to observe the creation of large-scale flux ropes without the latter leaving the box before MHD kinking can dissipate them, hence in our analysis this choice helps the physical understanding of reconnection dynamics.

In our work, we only considered the zero-guide-field case, and imposed no localized perturbation to initiate reconnection. This initial study paves the way for several follow-up works. In the future, it will be necessary to consider the presence of a guide field and/or the effect of initial perturbations to understand whether the results presented here hold in general. In particular, in most realistic scenarios, a guide field will most likely be present (albeit with wildly varying magnitudes depending on the context). In addition, given the ultrarelativistic energies reached by electrons and protons, we will need to consider radiation feedback onto the reconnection process. While these studies are beyond our current possibilities, they will be undertaken in the future to establish a comprehensive understanding of collisionless reconnection in realistic, ultramagnetized astrophysical scenarios.

\begin{acknowledgments}
F.B.\ would like to thank Dmitri Uzdensky and Lorenzo Sironi for useful discussions throughout the development of this work.
F.B.\ acknowledges support from the FED-tWIN programme (profile Prf-2020-004, project ``ENERGY'') issued by BELSPO, and from the FWO Junior Research Project G020224N granted by the Research Foundation -- Flanders (FWO). 
The resources and services used in this work were provided in part by the VSC (Flemish Supercomputer Center), funded by the Research Foundation - Flanders (FWO) and the Flemish Government. We acknowledge EuroCC Belgium for awarding this project access to the LUMI supercomputer, owned by the EuroHPC Joint Undertaking, hosted by CSC (Finland) and the LUMI consortium.
\end{acknowledgments}

\FloatBarrier
\bibliographystyle{aasjournal.bst}

{}

\end{document}